\documentclass[fleqn,usenatbib]{mnras}

\usepackage[T1]{fontenc}
\usepackage[normalem]{ulem}
\DeclareRobustCommand{\VAN}[3]{#2}
\let\VANthebibliography\thebibliography
\def\thebibliography{\DeclareRobustCommand{\VAN}[3]{##3}\VANthebibliography}

\usepackage{graphicx}	% Including figure files
\usepackage{amssymb}	% Extra maths symbols

\usepackage{amsmath}	% Advanced maths commands

\usepackage{newtxtext,newtxmath}
\title[X-ray spectropolarimetric models of accretion]{Spectral and polarization properties of reflected X-ray emission from black hole accretion discs}

\author[J. Podgorn\'y et al.]{
J. Podgorn\'y$^{1,2,3}$\thanks{E-mail: jakub.podgorny@astro.unistra.fr}, 
M. Dov{\v{c}}iak$^{2}$, %\thanks{E-mail: michal.dovciak@asu.cas.cz (MD)},
F. Marin$^{1}$, %\thanks{E-mail: frederic.marin@astro.unistra.fr (FM)},
R. Goosmann$^{1}$ %\thanks{E-mail: goosmann@astro.unistra.fr (RG)}
and A. R\'o\.za\'nska$^{4}$%\thanks{E-mail: agata@camk.edu.pl (AR)}
\\
$^{1}$Universit\'e de Strasbourg, CNRS, Observatoire Astronomique de Strasbourg, UMR 7550, F-67000 Strasbourg, France\\
$^{2}$Astronomical Institute, Academy of Sciences of the Czech Republic, Bo{\v{c}}n\'i II, CZ-14131 Prague, Czech Republic\\
$^{3}$Astronomical Institute, Charles University, V Hole{\v{s}}ovi{\v{c}}k\'ach 2, CZ-18000 Prague, Czech Republic\\
$^{4}$Nicolaus Copernicus Astronomical Center, Polish Academy of Sciences, Bartycka 18, 00-716 Warsaw, Poland
}

\date{Accepted 2021 December 16. Received 2021 November 20; in original form 2021 October 8}

\pubyear{2022}

\begin{document}
\label{firstpage}
\pagerange{\pageref{firstpage}--\pageref{lastpage}}
\maketitle

\begin{abstract}
X-ray polarimetric missions planned for this decade will significantly enhance our knowledge of compact accreting sources. Observations of the X-ray polarization signal from active galactic nuclei (AGNs) or X-ray binary systems (XRBs) will bring new means to study inner accretion flow in these objects that, together with currently used spectroscopic and timing techniques, will help us to determine better their properties, such as their inclination, orientation, shape, and size of their corona as well as the black hole spin. In this work, we present a yet missing piece in the global polarization models of black hole accretion discs. We compute the reflected X-ray emission from the disc in a local co-moving frame using (1) the radiative transfer code {\tt TITAN} to obtain the ionization structure of the disc and (2) the Monte Carlo code {\tt STOKES} that incorporates the physics of absorption, re-emission, and Compton scattering to produce a complete spectropolarimetric output. We present the final Stokes parameters $I$, $Q$ and $U$ for a set of photon-indices of the incident primary power-law radiation, the disc ionization parameters, incident and emission angles, for three independent polarization states of the incident coronal X-ray photons with a sufficient resolution in energy to allow for sharp discussion of spectral and polarization properties. We show that the spectral component matches well literature predictions. The polarization degree and angle are in agreement with analytical approximations previously appearing in reflection models and we demonstrate that the polarized reflected X-ray emission can be, locally, quite large in the 2--12 keV band.

\end{abstract}


%%%%%%%%%%%%%%%%%%%%%%%%%%%%%%%%%%%%%%%%%%%%%%%%%%

%%%%%%%%%%%%%%%%% BODY OF PAPER %%%%%%%%%%%%%%%%%%

\section{Introduction}

Spectroscopic X-ray observations from the past decades has led to a better understanding of compact accreting sources, such as black holes (BHs) or neutron stars (NSs). These objects gradually attract matter from their surroundings which results in the formation of an accretion disc that produces tremendous amounts of light, in particular in the X-ray band \cite[see e.g.][]{Seward2010}. Although these high-energy processes are considered quite omnipresent both in our Galaxy (in X-ray binary systems –- XRBs) and in the rest of the Universe (in extragalactic sources such as active galactic nuclei –- AGNs), a full understanding of their observed X-ray spectra is until today a difficult task. Considering mainly inner workings in radio quiet systems, i.e. excluding the presence of jets and the outer components of accreting systems, the central engine itself still poses important challenges to radiative transfer modelling in the vicinity
of strongly gravitating stellar-mass objects and supermassive BHs \citep{Antonucci1993,Done2007,Trumper2008,Seward2010,Abramowicz2013,Netzer2015}.

Re-opening of the X-ray polarimetric sky with the new generation, broad-band X-ray polarimeter on board of the IXPE mission \citep{Weisskopf2013, Weisskopf2016}, due to be launched by the end of 2021, or the eXTP mission \citep{Zhang2016, Zhang2019}, due to be launched in the second half of 2020s, will help us to solve some of the current problems. The forthcoming polarimetric measurements of X-ray sources urge for (re)investigation of polarization models, in addition to plain spectral modelling. For example, the energy dependence of X-ray polarization can provide an estimate of the BH spin \citep{Connors1977,Stark1977,Connors1980, Dovciak2008, Schnittman2009, Li2009, Taverna2020, Taverna2020b}, in addition to the well-known spectroscopic techniques (using either the iron K$\alpha$ line profile or the thermal disc continuum emission) or timing (kHz QPOs) techniques \cite[see][and references therein]{Reynolds2019}. The sensitivity of polarization to the geometry of sources allows to investigate also other global parameters of the system, such as the system overall inclination and orientation on the sky of the observer \citep{Dovciak2004, Dovciak2008, Li2009, Dovciak2011, Marin2014, Marin2016}. Thus, together with the recent rise of observational X-ray polarimetry, detailed computations of X-ray emission from compact accreting sources could provide a big step towards correct data interpretation in this energy band and towards a better understanding of the high-energy Universe.

In this article, we re-examine the spectral and polarization properties of the reflected X-ray emission from AGN discs. To do so, we construct a numerical model that follows the general consensus of X-ray emission: a power-law X-ray continuum component representing emission from a hot patch of gas -– the corona -– situated above the accretion disc \citep{Haardt1993, Haardt1993b, Dove1997, Krolik1999, Seward2010}. Such $\sim E^{-\Gamma}$, power-law emission, where $E$ is photon energy and $\Gamma$ is photon index, is due to the reprocessing of thermal radiation by Compton up-scattering processes. This primary radiation is then partially reflected by matter in the accretion disc mostly via Compton down-scattering and thus forms a secondary X-ray component that may reach the observer with distinct spectral and polarization signatures. In the case of AGNs, where the disc temperatures are not expected to produce significant amount of thermal radiation in the X-ray band, especially in the energy band above $1 \textrm{ keV}$ \citep{Shakura1973, Novikov1973, Reynolds2003, Abramowicz2013, Compere2017, Kubota2018}, the reflection tables, computed in a local fluid frame, should already realistically complement the models of primary radiation. The primary and reflected emission altogether with general-relativistic effects already comprise realistic total X-ray models of AGNs observed from infinity with the absence of jets and the absence of further radiation reprocessing via outer components, such as the broad-line regions (BLRs), narrow-line regions (NLRs), or a dusty torus \citep[see e.g.][]{Matt1993,Fabian2000,Seward2010,Netzer2015,Kubota2018}. Although in the case of XRBs, it is necessary to account for an extra thermal component of the X-ray radiation originating from the disc \citep{Rozanska2011}, the local reflection models presented here may also serve as a basic constituent for production of the total X-ray emission models of accreting stellar-mass compact objects.

Numerical attempts to compute the reflected X-ray spectrum under a power-law illumination of a disc were done most notably in the past with the codes {\tt PEXRAV} and {\tt PEXRIV} \citep{Magdziarz1995}, and {\tt REFLIONX} model \citep{Ross1993, Ross1999, Ross2005}, and more recently with the {\tt XILLVER} model \citep{Garcia2010, Garcia2011, Garcia2013}), including a detail computation of ionization structure of the disc by the code {\tt XSTAR} \citep{Kallman2001}. All the above computations assumed the reflection from a constant density slab, while stratified atmosphere in hydrostatic equilibrium was also calculated with {\tt TITAN} and {\tt NOAR} code \citep{Rozanska2002,Dumont2003}, and with {\tt ATM24} code \citep{rozanska2008,Rozanska2011,vincent2016}. The most precise X-ray polarization reflection models until today consist of only analytical approximations of scattering processes and are scarce. Attempts that were able to provide estimates on the polarization degree and angle of the reflected radiation from ionized accretion disc were presented in \cite{Dovciak2004, Schnittman2009, Dovciak2011}, assuming the Chandrasekhar’s formulae \citep{Chandrasekhar1960} for scattering. In this article, we present the first attempt to numerically simulate both the spectral and polarization outcome of local X-ray reflection with respect to various incident and emergent angles, ionization parameters, photon-indices and polarization states of the primary radiation. In order to achieve this, we used the Monte Carlo radiative transfer code {\tt STOKES} \citep{Goosmann2007,Marin2012,Marin2015,Marin2018} and pre-computed the ionization structure of the disc with the photoionization code {\tt TITAN} \citep{Dumont2003}, suitable for hot optically thick media.

The plain spectral output of our Monte Carlo approach serves as a completely novel reflection spectral model that may be discussed with respect to the previous attempts by codes that aim for the same with different methods, typically solving the equations of radiative transfer. The polarization degree and angle computed upon the spectral results with {\tt STOKES} then uniquely complement and improve the existing models with Chandrasekhar’s formulae, which may also
confirm correctness of the new simulation, if we demonstrate similar behaviour with respect to the main model parameters. Although the results presented in this paper do not represent full high-resolution tables, with the grid that was performed it is already possible to discuss important spectropolarimetric properties of the locally reflected radiation and to draw reasonable conclusions on the {\tt TITAN} and {\tt STOKES} codes’ performance with respect to the other approaches mentioned.

Such analysis may then serve as a basis to the total spectropolarimetric X-ray models of AGNs that the authors plan to achieve in the near future using these local reflection computations. Contrarily to the mildly ($\lesssim 10$ per cent) polarized primary radiation \citep{Beheshtipour2017, Tamborra2018, BeheshtipourThesis}, the total polarization signal is expected to be significantly enhanced due to additional polarization signatures emerging from Compton scattering inside the illuminated disc \citep{Schnittman2009, Dovciak2011}. The total simulated spectra, possibly including also the returning radiation \citep{Schnittman2010, Taverna2020}, will then provide direct answer to the total polarization enhancement by reflection, which is physically well-reasoned to emerge, especially at hard X-rays. If the polarization induced by reflection pro v es to be important, the presented tables will also serve as an estimate tool for observational times needed by the upcoming X-ray polarimetric missions for the faint AGN sources in the context of highly photon-demanding polarimetry \citep{Fabiani2014}.

Recently, a similar attempt to compute polarization of thermal disc emission in stellar-mass systems including absorption effects was performed in \cite{Taverna2020b} with the codes {\tt CLOUDY} \citep{Ferland2013, Ferland2017} and {\tt STOKES}. Therefore, this work on reflection tables could also complement the ongoing research for XRBs, which are also on the target lists of IXPE or eXTP missions and where the polarization induced by reflection may also play an important role. In the near future, the presented local computations could serve as a foundation stone for physically consistent X-ray polarimetric data fitting of AGN or XRB sources.

The structure of this paper is as follows: Section \ref{section2} provides an overview of the physical model assumed, the numerical techniques performed, and the parametric grid that was used. In Section \ref{section3}, we present the spectral and polarimetric properties of the new model. Strengths and weaknesses of the novel model and comparisons with similar approaches are further discussed in Section \ref{section4}. In Section \ref{section5}, we conclude our analysis and lay the ground for more extensive, future research.

\section{The model and numerical implementation}\label{section2}

\subsection{Physical model of the BH and its accreting structure}

The inner disc of a compact accreting source can be locally assumed as a plane-parallel slab with an electron scattering-dominated atmosphere \citep{Shakura1973, Novikov1973}. In the local co-moving frame of the disc, we approximate the otherwise geometrically thin disc as a semi-infinite optically thick slab with a constant density $n_\mathrm{H} = 10^{15} \textrm{ cm}^{-3}$. We stress that for acquiring accurate reflection spectra (for e.g. spectral line fitting), it is necessary to treat the stratified disc surface in full hydrostatic equilibrium, which causes thermal ionization instability and narrow two-phase zones in the medium between the disc and the corona \citep[see e.g.][]{Begelman1983,Rozanska1996}, even though the detailed geometry and boundary conditions of the illuminated disc atmosphere are unknown \citep{Ballantyne2001, Ross2005, Ross2007}. It has been shown in the literature that constant-density models vary from those of hydrostatic equilibrium in emission line shapes, including the iron line complex around $6$--$7 \textrm{ keV}$, as well as in the continuum \citep{Nayakshin2000, Nayakshin2001, Pequignot2001, Ballantyne2001, Rozanska2002, Dumont2002, Ross2007, rozanska2008, Rozanska2011, vincent2016}. The direct comparisons are not easy, since the ionization parameter has no meaning in the case of the hydrostatic equilibrium and it is also hard to define an average density \citep{Rozanska2002}. Despite the ongoing discussion, our aim in this paper is to compute the polarization properties of the reflection in the same scenario that is usually used for interpreting the spectra of AGNs and XRBs. In the observed X-ray spectra of these objects, the reflection component is usually fitted with the models assuming a constant density slab. The difference between the reflection spectra for the constant density and hydrostatic equilibrium slabs would then be manifested in our polarization computations mainly through the polarization degree energy dependence, where the amount and shape of depolarization due to spectral lines would change.

We represent the proximity and luminosity of the primary power-law source by an ionization parameter in $[\textrm{erg} \, \textrm{cm} \, \textrm{s}^{-1}]$ \citep[see e.g.][]{tarter1969}
\begin{equation}\label{xi}
	\xi = \dfrac{4 \pi \int F_E (r) dE}{n_\mathrm{H}} \textrm{ ,}
\end{equation}
where $F_E(r) \sim E^{-\Gamma+1}$ is the radiation energy flux locally received on the surface of the disc, and the bounds of the integral follow the chosen high and low energetic cut-offs (see Section \ref{nummodels} for our selection). Cut-off on the hard-energy tail is motivated by inability of coronal electrons to gain speeds that would result in such hard emission in the estimated coronal conditions of $T \approx 10^9 \textrm{ K}$. Low-energy cut-off is given by the energy of the seed thermal photons illuminating the corona, i.e. there are only few scattered photons with lower energy. For the case of our computations with {\tt STOKES}, we will denote the ionization parameter as $\xi_\mathrm{S}$. The range of the $\xi_\mathrm{S}$ and $\Gamma$ parameters studied was motivated by observational and physical prospects as well as by easily achievable comparison with previously published models. We adopted the typical iron solar abundance from \cite{Asplund2005} with $A_\mathrm{Fe} = 1.0$ (notation relative to the solar value), neglecting the presence of dust.

Our objective is first to pre-compute the vertically stratified ionization structure of the disc using isotropic illumination and then to continue with the 3D Monte Carlo polarization multiple-scattering simulation, incorporating appropriate line and continuum processes that take place in the disc. As we primarily focus on discs around supermassive BHs, we neglect in our X-ray band study the illumination from the thermal disc below the atmosphere. We neglect any possible primary radiation impact from the opposite side of the disc or self-irradiation by the disc in strong-gravity regime, and retain only the reflection effect of the power-law X-rays from the corona on top, which is expected to be Comptonized thermal radiation \citep{Haardt1993, Haardt1993b, Schnittman2010}.

In order to efficiently use the local reflection tables in total emission computations in future works and to be able to discuss the effect of primary polarization degree and angle on the reflected emission, we assumed three distinct polarization states of the primary radiation for our runs: (a) unpolarized light with the normalized Stokes parameters\footnote{The Stokes parameters $I$, $Q$, $U$ in all of this work correspond to their standard definitions in e.g. \cite{Chandrasekhar1960} and the orientation convention in the polarization plane is further clarified in Section \ref{section3} with the definition of polarization angle.} $Q_\mathrm{P} = 0$ and $U_\mathrm{P} = 0$, (b) horizontally polarized light with $Q_\mathrm{P} = -1$ and $U_\mathrm{P} = 0$, (c) diagonally polarized light with $Q_\mathrm{P} = 0$ and $U_\mathrm{P} = 1$. The illumination and reflection from the slab was then studied for all possible incident $\mu_\mathrm{i}$ angles and emergent $\mu_\mathrm{e}$, $\Phi_\mathrm{e}$ angles. We denote $\mu_\mathrm{i}$ as the cosine of an angle measured from the disc’s normal, and $\mu_\mathrm{e}$ as the cosine of an angle measured from the disc’s normal. We denote $\Phi_\mathrm{e}$ as the relative azimuthal angle measured counterclockwise at the disc plane between the emission and the incident ray directions projected to the disc’s surface [see e.g. fig. 1 in \cite{Dovciak2011} for a sketch of the different local reflection angles].

\subsection{Numerical models}\label{nummodels}

\subsubsection{TITAN}

{\tt TITAN} \citep{Dumont2003} is a code designed for computing the radiative transfer together with the ionization structure of the optically thick media. In our application, we used the latest version of the code updated for many new atomic transitions (see \citet{adhikari2015,goosmann2016}). We use it to pre-compute the ionization structure, i.e. the fractional abundance of each element in different states of ionization inside a plane–parallel slab of gas in thermal and ionization equilibrium, isotropically illuminated by a given power-law primary spectrum from one side. The photoionization radiative transfer is done using the \textit{accelerated lambda iteration} (ALI) method, while the ionization balance is computed in the full non-LTE approach, with iterations until the column depth $N_{\rm H} = 10^{25} \ \textrm{cm}^{-2}$ that corresponds to the Thomson optical depth $\tau \sim 7$. {\tt TITAN} solves the energy balance, the ionization, and the statistical equilibria, the transfer equations in a plane–parallel geometry, for the lines and continuum up to 26 keV, which is enough for the required output on line properties. For the purpose of this paper, we assume constant gas density over the slab with $n_{\rm H}=10^{15}$~cm$^{-3}$. The primary photon-flux in {\tt TITAN} has two exponential cut-offs and follows
\begin{equation}
	N(E) = N_0E^{-\Gamma}e^{-\frac{E}{E_\mathrm{T,c}}}e^{-\frac{E_\mathrm{T,0}}{E}}
\end{equation}
with much larger energy range between $E_\mathrm{T,0} = 0.03 \textrm{ keV}$, and  $E_\mathrm{T,c} = 300 \textrm{ keV}$. The output of {\tt TITAN} is then processed by a specific C++ script that produces a suitable input file for {\tt STOKES}, containing all the ionization structure information.

\subsubsection{STOKES}

{\tt STOKES} \citep{Goosmann2007,Marin2012,Marin2015,Marin2018} is a Monte Carlo radiative transfer code that was originally developed to solve near-IR to UV radiative transfer in 3D geometric shapes where scattering is the main source of opacity (as in, e.g. AGNs). In our case, we used the upgraded version \textit{2.32} of {\tt STOKES}, optimized for modelling of X-ray radiation, enabling reproduction of all 4 Stokes parameters $I$, $Q$, $U$, $V$ (in all the processes that we examine $V = 0$) for arbitrary incident polarization and ($\mu_\mathrm{e}$, $\Phi_\mathrm{e}$) bins at the surface ($z = 0$) of the pre-computed plane–parallel slab with 50 vertical layers. The emission region, from which the photons are injected inside the slab, is located at $z = 0$ and we prescribe also an incident angle $\mu_\mathrm{i}$ grid as one of the {\tt STOKES} input parameters. No bulk motion is considered for the emission region. We use a primary photon-flux with the three distinct polarization states according to
\begin{equation}\label{primary}
	N(E) = N_0E^{-\Gamma} \textrm{ ,}
\end{equation}
where the cut-offs are sharp at $E_\mathrm{min} = 10^{-1.1} \textrm{ keV}$ and $E_\mathrm{max} = 10^{2.4} \textrm{ keV}$. In order to guarantee similar amount of numeric noise across the energy range, we imposed that the photons are sampled equally in a logarithmically spaced grid in energy between $E_\mathrm{min}$ and $E_\mathrm{max}$ (350 bins with $\Delta \log E = 0.01$) with weight assignment to each photon according to the power-law distribution, for which the results are compensated at the output. The energy resolution was motivated by reasonable computational times on one hand and on the other by ability to provide reasonable spectro-polarimetric analysis with respect to other known models and with respect to the past and forthcoming X-ray missions’ resolution (IXPE, eXTP, XMM–Newton, ATHENA). The number of primary photons $N_\mathrm{tot}$ per one simulation run varied with $\xi_\mathrm{S}$, $\Gamma$, and $\mu_\mathrm{i}$, but was always $\geq 3.5 \times 10^8$ with the aim to reduce numerical noise as much as possible within reasonable computational times.

During the simulation, photons are followed along their trajectory and experience multiple interactions resulting in the line and continuum spectral and polarization features, such as Compton down-scattering, free–free interactions, photoelectric absorptions, etc. The {\tt STOKES} \textit{2.32} version used for our case did not account for (inverse) Compton up-scattering, nor synchrotron emission, which forms an acceptable compromise between complexity of the calculations and the energy range studied. A set of virtual detectors at the ($\mu_\mathrm{e}$, $\Phi_\mathrm{e}$) bins subsequently registers all the photons that were not absorbed within the multiple-scattering region. At each detector, the Stokes parameters of the collected photons are summed after having rotated the Stokes parameter reference frames around the detector line of sight, to match with the detector frame. An auxiliary routine {\tt ANALYZE} created by the authors of {\tt STOKES} is used to process the results into conveniently readable text files. In the end, we receive the Stokes parameters with respect to a parametric grid of $E$, $\xi_\mathrm{S}$, $\Gamma$, $\mu_\mathrm{i}$, $\mu_\mathrm{e}$, $\Phi_\mathrm{e}$ for the unpolarized, horizontally, and diagonally polarized states of the primary radiation.

\subsubsection{Resulting FITS tables}

In order to provide the computations compactly and in a well arranged way to the user, we decided to create and attach FITS files \citep{FITS} containing our results in reduced energy range and for 11 out of 20 originally computed $\mu_\mathrm{e}$ values between $\mu_\mathrm{e,min} = 0.025$ and $\mu_\mathrm{e,max} = 0.975$ with $\Delta \mu_\mathrm{e} = 0.05$. We preserve all the computed $\mu_\mathrm{i}$ with $\Delta \mu_\mathrm{i} = 0.1$ and $\Phi_\mathrm{e}$ with $\Delta \Phi_\mathrm{e} = 15^{\circ}$. The output Stokes parameters $I$, $Q$ and $U$, here below commonly denoted as $S$, were stored to the FITS files with the following normalization
\begin{equation}\label{norm1}
	N(E;\mu_\mathrm{e}, \mu_\mathrm{i}, \Phi_\mathrm{e}) = \dfrac{A \xi_\mathrm{S} n_\mathrm{H}}{4 \pi e \times 10^{10}}\dfrac{S(E; \mu_\mathrm{e}, \mu_\mathrm{i}, \Phi_\mathrm{e})}{N_\mathrm{tot} \Delta \mu_\mathrm{e} \Delta \Phi_\mathrm{e}} \textrm{ ,}
\end{equation}
where $e$ is the elementary charge in coulombs necessary for a unit conversion and
\begin{equation}\label{norm_factor}
	A = 
	\begin{cases}
		\dfrac{(E_\mathrm{max}^{1-\Gamma}-E_\mathrm{min}^{1-\Gamma})(2-\Gamma)}{(1-\Gamma)(E_\mathrm{max}^{2-\Gamma}-E_\mathrm{min}^{2-\Gamma})} \ , & \Gamma \notin \{1,2\} \ , \\
		 \ln\left( \dfrac{E_\mathrm{max}}{E_\mathrm{min}}\right) \dfrac{2-\Gamma}{E_\mathrm{max}^{2-\Gamma}-E_\mathrm{min}^{2-\Gamma}}  \ , & \Gamma = 1 \ , \\
		 \dfrac{E_\mathrm{max}^{1-\Gamma}-E_\mathrm{min}^{1-\Gamma}}{1-\Gamma} \ln^{-1}\left( \dfrac{E_\mathrm{max}}{E_\mathrm{min}}\right)  \ , & \Gamma = 2 \ .
	\end{cases}
\end{equation}
In this way, the photon-flux is stored in the units of $[\textrm{counts} \, \textrm{cm}^{-2} \, \textrm{s}^{-1}]$. This normalization accounts for the differences in the cut-offs used in {\tt TITAN} and {\tt STOKES} codes for the primary radiation and takes into consideration the units in standard output of the codes. In total, nine FITS files were created for the combinations of three incident polarization states with the three linear ($I$, $Q$ and $U$) Stokes parameters. For storage convenience, we reduce the energy range to 300 bins between 0.1 and 100 keV. The adopted $\xi_\mathrm{S}$ grid varies with $\Gamma$, but in general $\xi_\mathrm{S}$ spans values between 1.1 and 39 225.5. For this reason, we store index values $\{1,2,...,11\}$ instead of real $\xi_\mathrm{S}$ and attach a separate extension at the end of the FITS file, which holds the real $\xi_\mathrm{S}$ values assigned to indices for each $\Gamma$. Apart from this last extension, the attached FITS files conform to the OGIP standard for {\tt XSPEC} \citep{OGIP, Arnaud1996} in this way.\footnote{We are currently computing with the same method the full OGIP standard FITS file tables for XSPEC with the ionization parameter grid that does not depend on $\Gamma$. These tables will be published in the near future.} Table \ref{FITSgrid} brings a complete overview of the attached FITS files and adopted parametric mesh.
\begin{table}
	\centering
	\caption{Description of the attached FITS tables representing the reduced local reflection tables computed by {\tt STOKES}.}
	\begin{tabular}{ll}
		\hline
		\hline
		Number of tables & 9 (horizontal, diagonal, and no incident polarisation \\
		&  versus $I$, $Q$, $U$ output normalized according to (\ref{norm1}))  \\
		Spectral units            & $[\textrm{counts} \, \textrm{cm}^{-2} \, \textrm{s}^{-1}]$        \\
		Energy range     & $0.1 \textrm{ keV}$ to $100 \textrm{ keV}$                       \\
		Energy binning      & 300 bins, $\Delta \log E = 0.01$                                             \\
		$\Gamma$            & {\{}$1.2, 1.4, 1.6, 1.8, 2.0, 2.2, 2.4, 2.6, 2.8, 3.0${\}}                                                             \\
		$\xi_\mathrm{S}$ index            & {\{}$1,2,3,4,5,6,7,8,9,10,11${\}}       \\
		$\mu_\mathrm{i}$            & {\{}$0.0,0.1,0.2,0.3,0.4,0.5,0.6,0.7,0.8,0.9,1.0${\}}                    \\
		$\mu_\mathrm{e}$            & {\{}$0.025,0.075,0.175,0.275,0.375,0.475,0.575,$   \\
		&
		\multicolumn{1}{l}{$\ \ \ 0.675,0.775,0.875,0.975${\}}}                        \\
		$\Phi_\mathrm{e}$           & {\{}$7.5^{\circ},22.5^{\circ},37.5^{\circ},52.5^{\circ},67.5^{\circ},82.5^{\circ},97.5^{\circ},$                               \\
		&
		\multicolumn{1}{l}{$\ \ \ 112.5^{\circ},127.5^{\circ},142.5^{\circ},157.5^{\circ},172.5^{\circ},187.5^{\circ},$}                                              \\
		& \multicolumn{1}{l}{$\ \ \ 202.5^{\circ},217.5^{\circ},232.5^{\circ},247.5^{\circ},262.5^{\circ},277.5^{\circ},$}                                \\
		& \multicolumn{1}{l}{$\ \ \ 292.5^{\circ},307.5^{\circ},322.5^{\circ},337.5^{\circ},352.5^{\circ}${\}}}         \\
		Extensions       & Primary Header  -- description of the tables                                             \\
		& `PARAMETERS’      -- parameter values                                                      \\
		& `ENERGIES’     -- low and high energetic bin edges                                      \\
		& `SPECTRA’  -- values of the Stokes parameters and \\
		&
		\multicolumn{1}{l}{corresponding model parametric values}           \\
		& `XI(GAMMA)’   -- conversion table to real $\xi_\mathrm{S}$ values\\
		&
		\multicolumn{1}{l}{depending on $\Gamma$ and $\xi_\mathrm{S}$ index} \\ \hline \hline
	\end{tabular}%
	\label{FITSgrid}
\end{table}

\section{Results}\label{section3}

In all the spectral figures presented below, we display results per keV and use the normalization (\ref{norm1}) or the following normalization in case of integrated {\tt STOKES} spectra in angular space ($N_{\mu_\mathrm{e}} = 20$ in our case)
\begin{equation}\label{norm2}
	N(E) = \dfrac{A\xi_\mathrm{S} n_\mathrm{H}}{N_{\mu_\mathrm{e}} 4 \pi^2 e \times 10^{10}} \sum_{\mu_\mathrm{e},\mu_\mathrm{i},\Phi_\mathrm{e}} \mu_\mathrm{i} \Delta\mu_\mathrm{i}  \dfrac{S(E; \mu_\mathrm{e}, \mu_\mathrm{i}, \Phi_\mathrm{e})}{N_\mathrm{tot}\Delta\mu_\mathrm{e}} \textrm{ ,}
\end{equation}
if not mentioned otherwise. The linear polarization fraction $p$ and the polarization angle $\Psi$ are obtained from the {\tt STOKES} output through the usual expressions
\begin{equation}\label{ppsidef}
	\begin{aligned}
		p &= \dfrac{\sqrt{Q^2+U^2}}{I} \\
		\Psi &= \dfrac{1}{2}\textrm{\space}\arctan_2\left(\dfrac{U}{Q}\right)   \textrm{ ,}
	\end{aligned}
\end{equation}
where $\arctan_2$ denotes the quadrant-preserving inverse of a tangent function and $\Psi = 0$ corresponds to a polarization vector oriented along the projected disc’s normal to the polarization plane. $\Psi$ increases in the counterclockwise direction from the observer’s point of view. Displays of polarization degree with energy have been averaged over the neighbouring 10 bins in energy [first $Q$ and $U$ in the output were averaged and then transformed via (\ref{ppsidef})] for the sake of numerical noise reduction but not to the detriment of physical information loss. But even this reduced resolution is far better than any X-ray polarimetric instrument planned for the IXPE \citep{Weisskopf2013, Weisskopf2016} or eXTP \citep{Zhang2016, Zhang2019} missions.

\subsection{Spectral properties}

\begin{figure}
	\includegraphics[width=\columnwidth]{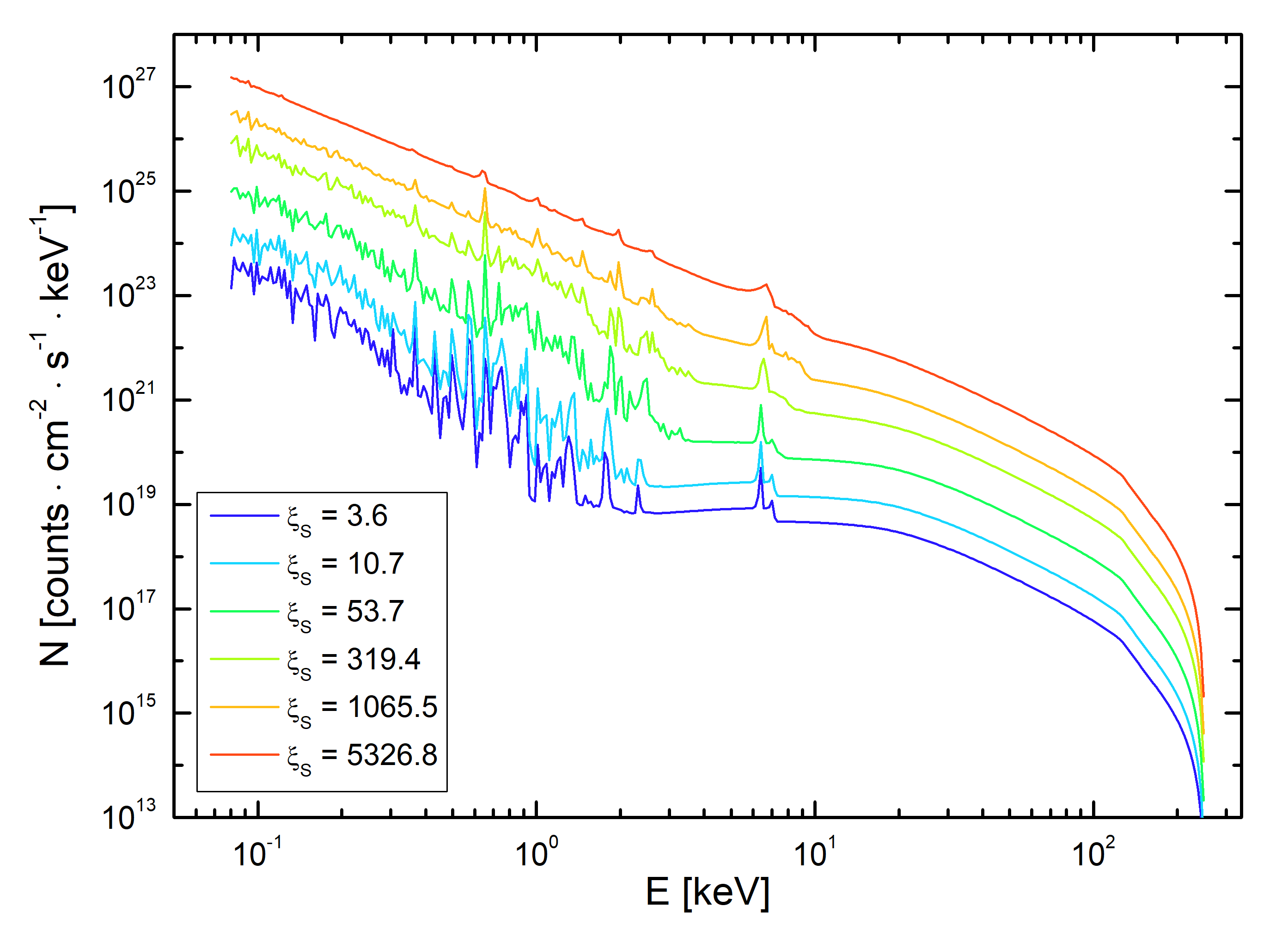}
	\caption{{\tt STOKES} spectral results integrated in $\mu_\mathrm{i}$, $\mu_\mathrm{e}$ and $\Phi_\mathrm{e}$ for the unpolarized primary radiation, $\Gamma = 2.2$. The color code corresponds to various $\xi_\mathrm{S}$.}
	\label{fig:spectrumxi}
\end{figure}

We first show the X-ray spectral output $N(E)$ obtained by {\tt STOKES} for initially unpolarized radiation (to be compared with {\tt XILLVER} results in Section \ref{section4}). Fig. \ref{fig:spectrumxi} shows spectral results integrated in angular space according to (\ref{norm2}), for various $\xi_\mathrm{S}$ and one $\Gamma$. All the spectra possess characteristic features \citep[see e.g.][]{Fabian2000, Remillard2006, Seward2010} such as the Compton hump at around 20 keV, a sharp decline at $E > 10^2 \textrm{ keV}$, an excess at soft energies, a forest of lines around 1 keV, and the most prominent Fe K$\alpha$ line at $6.4$--$7 \textrm{ keV}$, also confirmed by large number of observations from X-ray accreting sources \citep{Gottwald1995, Winter2009, Ng2010}. The expected result at energies above 10 keV was achieved in {\tt STOKES} , i.e. that Compton scattering, governed by the fairly constant Klein–Nishina cross-section with energy, is the dominant source of opacity at hard X-rays to the detriment of photoelectric absorption, which decays as $\sim E^{-3}$ \citep{Garcia2013}. The behaviour of $N(E)$ with $\xi_\mathrm{S}$ corresponds to the cases in literature (see e.g. fig. 2 in \cite{Fabian2000}). Emission lines are superimposed to the continuum but become completely yielded once the ionization and reflection capabilities of the disc become higher and the re-processed radiation begins to resemble the shape of the original incident spectrum. \cite{Fabian2000} discusses four distinct regimes of spectral variation with $\xi$ based on atomic properties and absorption edges involved, which is supported by {\tt STOKES} computations and the resulting Fig. \ref{fig:spectrumxi}.

A new result brought by our work is the role of incident polarization on the reflected spectral outcome. Although some X-ray polarization of the primary radiation is expected to be due to Comptonization of thermal radiation inside the corona, general estimates for the polarization degree of the primary source are rather low ($\lesssim 10\%$) according to the recent analysis in \cite{Beheshtipour2017, Tamborra2018, BeheshtipourThesis}. If some polarization was present, then, for symmetric reasons, vertically or horizontally polarized light should be dominant in case of extended coronal models \citep{Dabrowski2001,Niedzwiecki2008, Schnittman2010} at low heights. In case of lamp-post models \citep{Matt1991,Martocchia1996,Henri1997, Martocchia2000, Dovciak2004b, Miniutti2004} general-relativistic effects will rotate the polarization position angle along null geodesics from the corona towards the disc \citep{Connors1980} and the situation becomes more general for incident disc irradiation, i.e. any incident state of polarization is possible. Our computations, apart from the unpolarized case, assumed two extreme cases of initial $p = 100$ per cent polarization in order to estimate its possible effects in comparison with the completely unpolarized light and in order to test that the code adheres to the stated orientation conventions. Having appended these two initially polarized cases to our tables, it also allows for interpolation of reflection results for any input polarization state from this basis of the three computed independent polarization states, which will be necessary in future construction of global spectropolarimetric models. We aim to address any impact of light bending and other relativistic effects for a distant observer in our future works that will introduce the {\tt STOKES} local tables integrated over the accretion disc with some adopted global geometry.

\begin{figure}
	\includegraphics[width=\columnwidth]{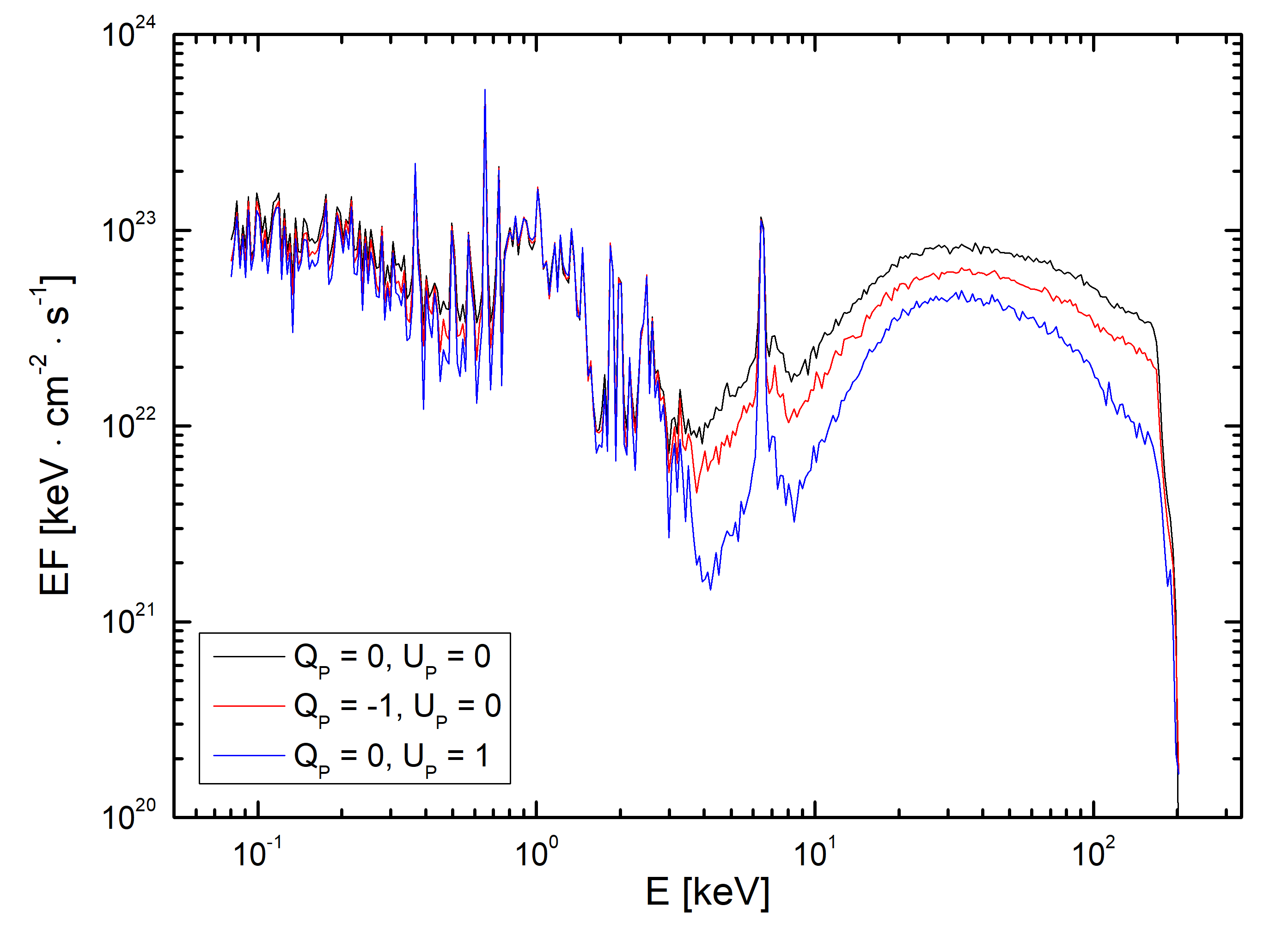}
	\caption{Directional spectrum obtained by {\tt STOKES} for $\mu_\mathrm{i} = 0.5$, $\mu_\mathrm{e} = 0.475$, $\Phi_\mathrm{e} = 292.5$, $\xi_\mathrm{S} = 143.4$ and $\Gamma = 2.0$. Three different primary polarizations are shown. Unpolarized light with $Q_\mathrm{P} = 0$ and $U_\mathrm{P} = 0$ (black), horizontally polarized light with $Q_\mathrm{P} = -1$ and $U_\mathrm{P} = 0$ (red), and diagonally polarized light with $Q_\mathrm{P} = 0$ and $U_\mathrm{P} = 1$ (blue). This example represents typical differences in spectra for distinct primary polarization in the local reflection tables.
	}
	\label{fig:ipolnonint}
\end{figure}

Typical effect of variable initial polarization on the local spectrum emitted in particular direction is shown on Fig. \ref{fig:ipolnonint}, which is about half order of magnitude in $EF = N E^2$ quantity (for $\Gamma = 2$ to abstract from plain slope effects) in the Compton hump region. We checked multiple selected incident and emission angles and their spectra under different initial polarization. Despite the role of multiple scattering that depolarizes the incident radiation and smears the differences at the spectral output, the damping of the spectral output expected for single-scattered photons in certain angular configurations was confirmed (initially 100 per cent polarized light should be strongly filtered, if oriented perpendicular to the plane of scattering, upon Compton scattering close to the $\Theta = 90^{\circ}$ scattering angle for single scatters). The amplitude of the spectral output was studied relative to other angular configurations. The maximum differences in the local reflected radiation for our three distinct initial polarization states achieve about an order of magnitude in $EF$ (for $\Gamma = 2$) in the Compton hump region, but may also completely vanish in the spectral output. Variation of initial polarization (for any angular configuration) behaves at the spectral output in the same manner for different $\xi_\mathrm{S}$ and $\Gamma$, as there is no prominent connection of $\xi_\mathrm{S}$ and $\Gamma$ parameters with this effect induced by the most common single-scattering angle in the re-processed radiation.

\begin{figure}
	\includegraphics[width=\columnwidth]{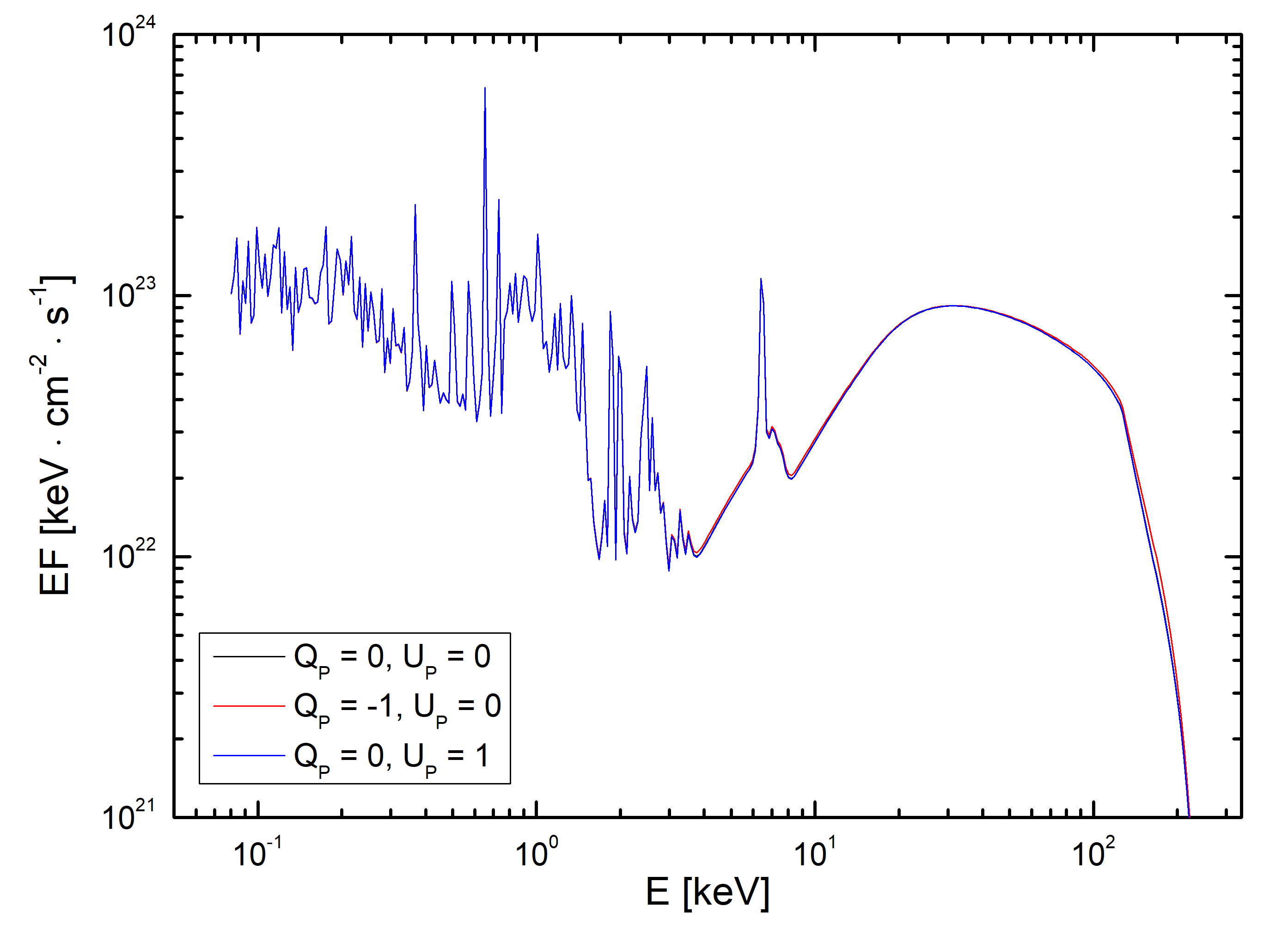}
	\caption{Spectrum obtained by {\tt STOKES} integrated in $\mu_\mathrm{i}$, $\mu_\mathrm{e}$ and $\Phi_\mathrm{e}$ for $\xi_\mathrm{S} = 143.4$ and $\Gamma = 2.0$. Three different primary polarization states are shown. Unpolarized light with $Q_\mathrm{P} = 0$ and $U_\mathrm{P} = 0$ (black), horizontally polarized light with $Q_\mathrm{P} = -1$ and $U_\mathrm{P} = 0$ (red), and diagonally polarized light with $Q_\mathrm{P} = 0$ and $U_\mathrm{P} = 1$ (blue).
	}
	\label{fig:ipolint}
\end{figure}

If we integrate over incident and emergent angles and display the spectrum for the three cases of initial polarization studied, the differences are inconspicuous, as displayed on Fig. \ref{fig:ipolint}. Assuming an integration over angles with uniform weights to all computed incident and emission angles according to (\ref{norm2}), we received average relative variation between the cases studied in the $EF = N E^2$ quantity of 0.3 per cent at $0.3$--$1$ keV, 2.5 per cent at 2--10 keV, and 2.0 per cent at 11--15 keV. The insensitivity of the direction-averaged spectra to the incident polarization are due to almost even spanning of all dominant single-scattering angles $\Theta$. In our future works, we plan to address the non-uniform integration expected for realistic strong-gravity environments.

Even though Fig. \ref{fig:ipolnonint} suggests that variable primary polarization may play a role for an amplitude of hard X-ray spectra in some strongly directional global GR configurations, one needs to realize that we show initially 100 per cent polarized light versus an unpolarized case, while coronal studies suggest low estimates on the primary polarization degree. Therefore, we still remain sceptical about any seeing of such spectral effects (either on the continuum or the Fe K$\alpha$ line) in realistic global models. However, if one brings attention to the \textit{polarization} output itself, i.e. the polarization degree and angle, which will be soon measured by the forthcoming X-ray polarimetric missions, the primary polarization will play a more significant role in the reflected light (see Section \ref{pprop}).

Strength of the reflected radiation with respect to the model parameter values can be obtained by displaying the total disc reflectivity, i.e. the quantity $N_\mathrm{r}/N_\mathrm{tot}$, where $N_\mathrm{r}$ denotes the number of photons detected in the {\tt STOKES} output for unpolarized primary radiation with a given $\mu_\mathrm{i}$, $\xi_\mathrm{S}$, and $\Gamma$, integrated in specific energy range and in all emission directions (i.e. in the entire $\mu_\mathrm{e}$ and $\Phi_\mathrm{e}$ angular space). This quantity will generally behave differently at soft X-rays, hard X-rays, and if integrated in the total energy range. None the less, we found that reflectivity slowly declines with increasing $\mu_\mathrm{i}$ in the soft, hard, and total energy range, regardless of $\xi_\mathrm{S}$ and $\Gamma$ choice. This tendency is natural, because more absorption is expected for high $\mu_\mathrm{i}$ as the incident photons will penetrate further below the disc’s surface in a more vertical incident ray direction. Since this tendency is similar for all parametric combinations, it is possible to further integrate disc reflectivity in $\mu_\mathrm{i}$, in order to obtain better statistics on its behaviour with respect to $\xi_\mathrm{S}$ and $\Gamma$ at different energy bands.

\begin{figure}
	\includegraphics[width=\columnwidth]{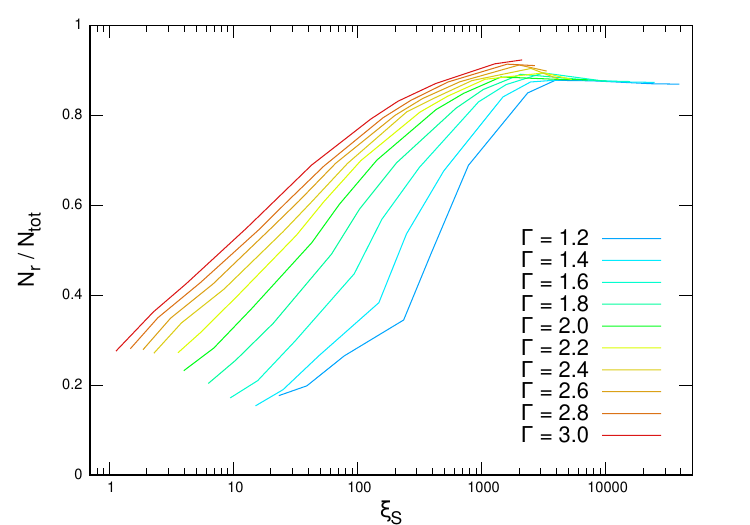}
	\caption{Total reflectivity obtained by {\tt STOKES} as it varies with $\xi_\mathrm{S}$ for different $\Gamma$, integrated over the total energy range and over all $\mu_\mathrm{i}$.}
	\label{fig:refltot}
\end{figure}

In Fig. \ref{fig:refltot}, we show the reflectivity with respect to $\xi_\mathrm{S}$ for various
$\Gamma$ cases, integrated over the full energy range and over all $\mu_\mathrm{i}$. The fact that reflectivity rises with increasing $\xi_\mathrm{S}$ is related to Fig. \ref{fig:spectrumxi}, which already confirmed that the disc tends to act more like a mirror of the primary illumination for higher ionization. Furthermore, harder X-ray illumination, i.e. lower $\Gamma$ parameter, yields lower total reflection, but for completely ionized discs there is no difference. The quantitative estimates of the total disc reflectivity may be important e.g. for thermal reverberation in the accretion disc, where it is assumed that a fraction of the illuminating coronal X-rays are absorbed and thus the disc is heated \citep[see, e.g.][]{Kammoun2019, Dovciak2021}.

\subsection{Polarization properties}\label{pprop}

The polarization properties of light will change when scattered of free electrons, or when line transitions are involved. Recombination lines come out unpolarized, because they statistically even out, even if single emission was polarized. Resonant lines and Compton scattering involves incoming photon with arbitrary polarization that is redistributed with polarization properties, which are ultimately governed by scattering angle, having the most effective polarization change for $\Theta \approx 90^{\circ}$. Therefore, in the first place and for single-scattered photons, geometrical discussion should focus upon the incident and emergent angles $\mu_\mathrm{i}$, $\mu_\mathrm{e}$, and $\Phi_\mathrm{e}$, which can altogether form right angles in numerous ways. Multiple scattering, which is more likely for larger $\mu_\mathrm{i}$, always distorts this dominant effect for single-scattered photons. Then, in the second place, we should reflect on energetic properties of possible (de)polarization and the energy-dependent cross-sections of these processes, including absorption of a photon. This is also coupled to geometry, as for different $\mu_\mathrm{i}$ various depths are likely to be reached and the cross-sections of line and continuum processes vary with the disc’s vertical stratification. Lastly, photons may change energy due to Compton down-scattering and one photon may encounter multiple kind of processes with different treatment of polarization during its journey through the medium. None the less, some distinct polarization features in locally reflected radiation still appear in the final output of the {\tt STOKES} simulations. They can be unambiguously explained by one or two origins.

\begin{figure}
	\includegraphics[width=\columnwidth]{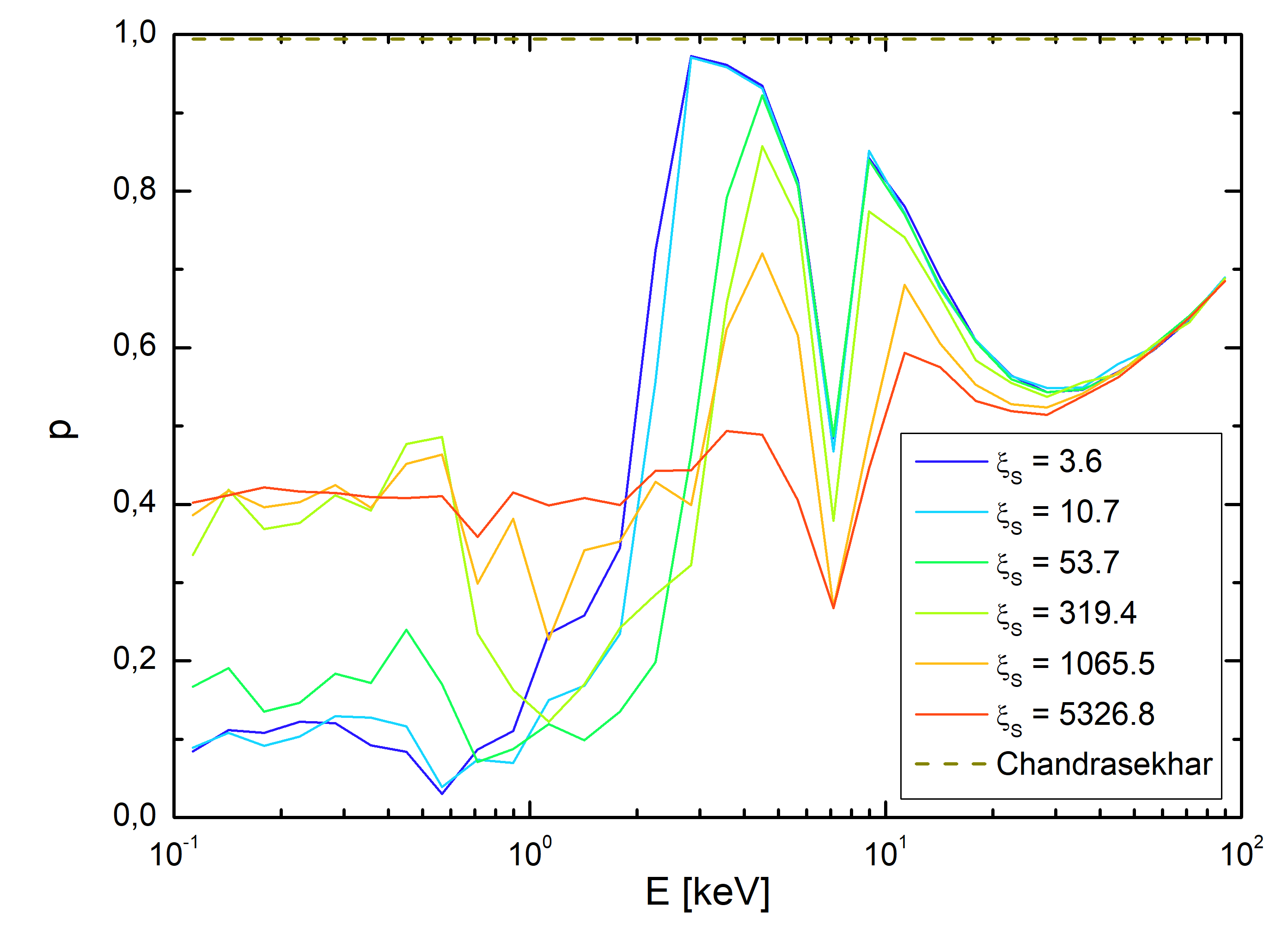}
	\caption{The polarisation degree obtained by {\tt STOKES} (solid lines) for the unpolarized primary radiation, $\mu_\mathrm{i} = 0.5$, $\mu_\mathrm{e} = 0.475$, $\Phi_\mathrm{e} = 292.5$, $\Gamma = 2.2$, and varying $\xi_\mathrm{S}$. Value for the single-scattering approximation by Chandrasekhar’s formulae for unpolarized primary radiation in this geometry is shown (dashed line, hardly visible on the top of the figure).}
	\label{fig:pxivar}
\end{figure}

\begin{figure}
	\includegraphics[width=\columnwidth]{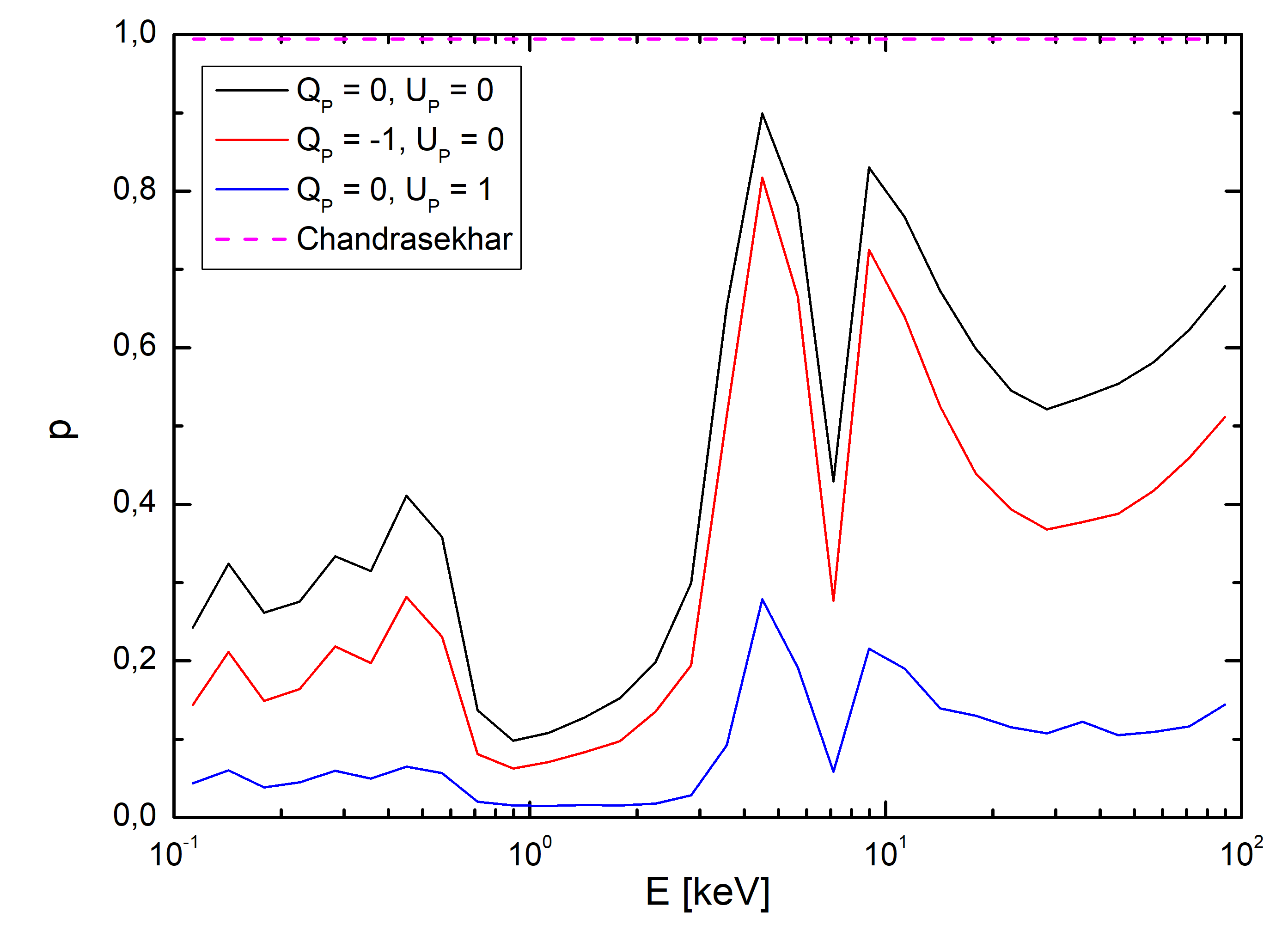}
	\caption{The polarization degree obtained by STOKES (solid lines) for {\tt STOKES} (solid lines) for $\mu_\mathrm{i} = 0.5$, $\mu_\mathrm{e} = 0.475$, $\Phi_\mathrm{e} = 292.5$, $\xi_\mathrm{S} = 143.4$ and $\Gamma = 2.0$. Three different primary polarizations are shown. Unpolarized light with $Q_\mathrm{P} = 0$ and $U_\mathrm{P} = 0$ (black), horizontally polarized light with $Q_\mathrm{P} = -1$ and $U_\mathrm{P} = 0$ (red), and diagonally polarized light with $Q_\mathrm{P} = 0$ and $U_\mathrm{P} = 1$ (blue). This example corresponds to the parametrization shown in Fig. \ref{fig:ipolnonint}. Value for the single-scattering approximation by Chandrasekhar’s formulae for unpolarized primary radiation in this geometry is shown (dashed line, hardly visible on the top of the figure).
	}
	\label{fig:ppol1}
\end{figure}

Regarding the polarization degree $p$, Fig. \ref{fig:pxivar} displays variation of $p(E)$ with $\xi_\mathrm{S}$ for unpolarized primary radiation and fixed generic set of remaining parameters. Fig. \ref{fig:ppol1} displays also the non-integrated $p(E)$ values for a similar set of parameters, but with varying initial polarization. Fig. \ref{fig:ppol1} uses the same set of model parameters as in Fig. \ref{fig:ipolnonint} and demonstrates higher relative impact of incident polarization conditions on $p$ than on the spectral output. It was already clear from Figs \ref{fig:ipolnonint} and \ref{fig:ipolint} that, at the soft part of the spectrum below 3 keV, where most spectral lines are present, the differences with respect to the initial polarization states are minor. Figs \ref{fig:pxivar} and \ref{fig:ppol1} confirm that spectral lines act destructively on any non-zero incident polarization and that the lines manifest through low $p$ at soft X-rays, and with a low dip in $p$ at the Fe K$\alpha$ line at 6.4--7 keV. Although some polarization would be expected from resonant lines, they do not appear dominant in the output, and may be also impacted by multiple scattering, which distorts any clear single scattering contribution on overall polarization state.

A tendency of rising polarization degree towards higher energies is visible, as less lines are present in the hard X-rays and pure Compton recoil (which is a dominant process for polarization emergence in our set-up) prevails. A large dip similar to the mirrored Compton hump shape in $N(E)$ is always present around 20 keV. This is expected, as multiple scattering happens in the Compton hump energy band, to which the photons from the primary radiation are largely redistributed
in energy. Hence, a decrease in polarization degree emerges over a generally large polarization degree in the hard X-rays, which are not affected by spectral lines. On Fig. \ref{fig:pxivar}, strong variations with $\xi_\mathrm{S}$ are less detectable in the high-energy tail, as none of the Stokes parameters varies with $\xi_\mathrm{S}$ there. In the low-energy tail, polarization rises with increasing $\xi_\mathrm{S}$ due to gradually less spectral lines being present (see Fig. \ref{fig:spectrumxi}). In the soft energy band, regarding the definition of $p$ (\ref{ppsidef}), the rise of intensity with $\xi_\mathrm{S}$ is compensated by even larger rise in $Q$ and $U$, while the slopes originating in primary power-law cancel out. Antagonistic trend around 10 keV cannot be explained by spectral lines presence and is due to natural rise of intensity (still observable on Fig. \ref{fig:spectrumxi} to the detriment of $Q$ and $U$ increment at these energies). The fact that absorption of photons (see Fig. \ref{fig:spectrumxi}) may result in enhanced polarization degree was already discussed in \cite{Taverna2020b}.

\begin{figure}
	\includegraphics[width=\columnwidth]{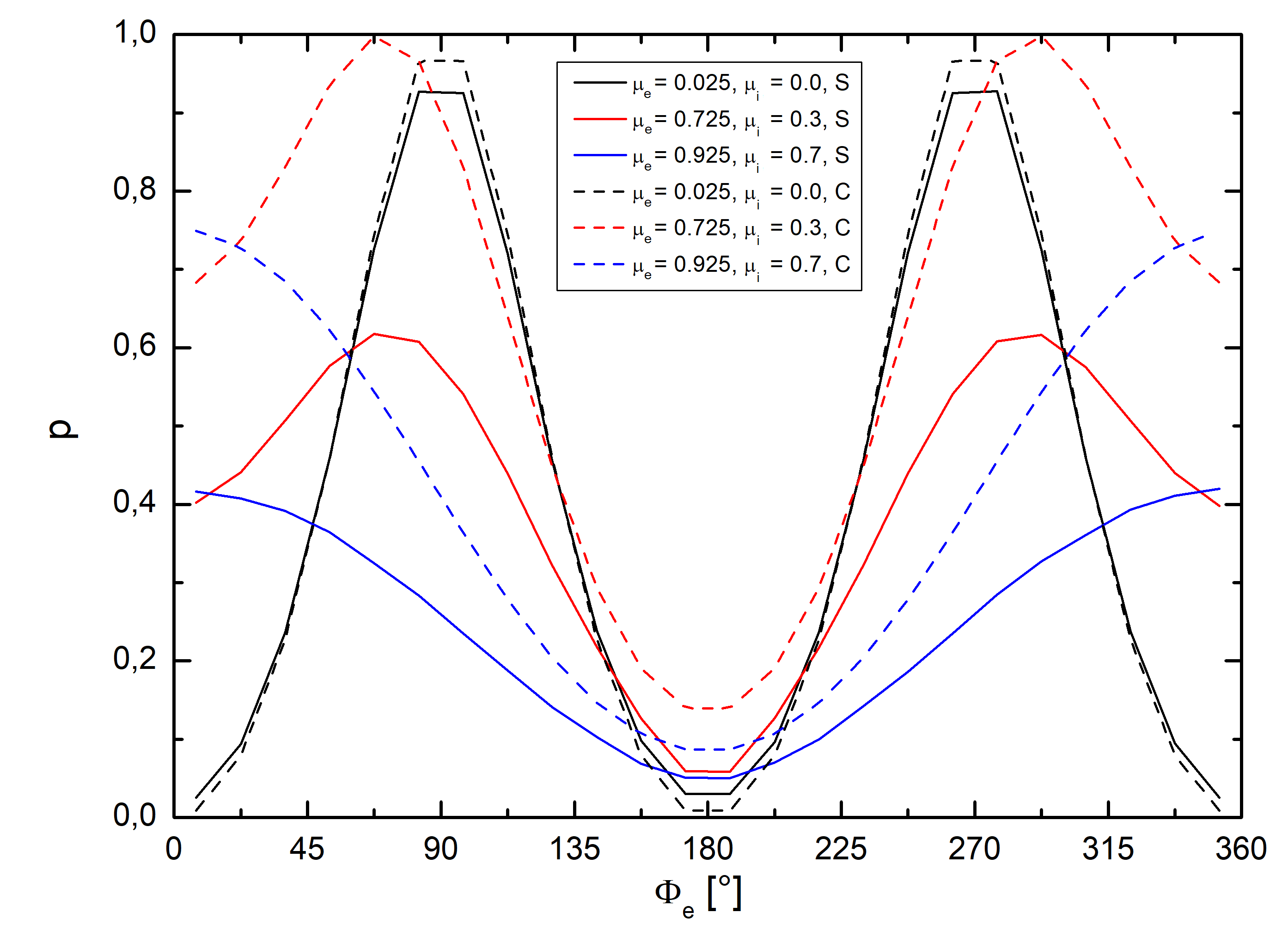}
	\caption{The polarisation degree versus azimuthal emission angle obtained by {\tt STOKES} (solid lines) integrated at $E > 10 \textrm{ keV}$ for unpolarized primary radiation, $\xi_\mathrm{S} = 106.5$, $\Gamma = 2.2$, and various $\mu_\mathrm{i}$, $\mu_\mathrm{e}$ combinations. Values for the single-scattering approximation by Chandrasekhar's formulae for unpolarized primary radiation in these geometries are shown (dashed lines).}
	\label{fig:pphivar}
\end{figure}

In order to study further effects with more clarity, we averaged $p$ over $E > 10 \textrm{ keV}$, where Compton scattering is by far the most important re-processing mechanism, and selected generic $\Gamma$ and $\xi_\mathrm{S}$. $p(\Phi_\mathrm{e})$ is displayed on Fig. \ref{fig:pphivar} for various $\mu_\mathrm{i}$ and $\mu_\mathrm{e}$ values. In this way, we may study the angular behaviour of polarization output with respect to the dominant single-scattering angle in {\tt STOKES} simulations and compare them to the Chandrasekhar’s formulae \citep{Chandrasekhar1960} for Rayleigh single scattering of unpolarized incident radiation inside a horizontal slab. These analytical values were already implemented to the previously mentioned graphs of $p$, but with little importance, as their values were overly high for the selected geometries, although it can be concluded that unpolarized incident radiation reached these idealized values around 3 keV. Fig. \ref{fig:pphivar} clearly displays the code’s results with respect to the analytical approach in the reflection scenario. {\tt STOKES} treats the scattering angles (via $\mu_\mathrm{i}$, $\mu_\mathrm{e}$ and $\Phi_\mathrm{e}$, which together set the dominant single-scattering angle) with precision and the polarization degree results are bilaterally symmetric in $\Phi_\mathrm{e}$.

The presence of multiple scattering and general crudeness of the Chandrasekhar’s approximation, unsuitable for reflection processes on the accretion discs of AGNs (further discussed in Section \ref{section4}), are the reasons for the lack of $p$ in the {\tt STOKES} semirealistic simulation behind the idealized analytical values. To further prove the effect of multiple scatterings, we put a condition in the {\tt STOKES} routines for only one electron scattering event per photon to mimic the single-scattering analytical approach in the Monte Carlo code. We tested two examples on Fig. \ref{fig:pphivar}, where the dashed and solid lines, i.e. the analytical and numerical approach, differ significantly: (a) $\Phi_\mathrm{e} = 67.5^{\circ}$, $\mu_\mathrm{e} = 0.725$, $\mu_\mathrm{i} = 0.3$, the red lines, and (b) $\Phi_\mathrm{e} = 67.5^{\circ}$, $\mu_\mathrm{e} = 0.925$, $\mu_\mathrm{i} = 0.7$, the blue lines; rest of the parameters remain as for Fig. \ref{fig:pphivar}. The average polarization degree above 10 keV in this approach was (a) 98.08 per cent (for the Chandrasekhar’s single-scattering approximation this is 99.77 per cent, for the {\tt STOKES} full multiple-scattering simulation this is 61.74 per cent) and (b) 53.12 per cent (for the Chandrasekhar’s single-scattering approximation this is 54.37 per cent, for the {\tt STOKES} full multiple-scattering simulation this is 32.50 per cent).

The polarization angle tends to oscillate around a constant value at all energies in the spectral lines, i.e. these oscillations are rather visible at the soft X-rays for lower $\xi_S$ values, as far as our adopted energy resolution allows to investigate. Apart from degenerate coordinate cases (because of presence of the so-called critical point in the definition of polarization angle discussed in, e.g. \cite{Dovciak2011}, when the polarization plane is perpendicular to the disc’s normal) or low polarization degree, where the most frequent value is ambiguous due to low statistics, the polarization angle of the continuum is otherwise constant in energy (apparent when one e.g. averages over the 10 neighbouring energy bins), because $Q$ and $U$ tends to behave in the same manner with $E$. This implies independence of the polarization angle of the continuum on $\Gamma$, and the independence of the polarization angle of the continuum on $\xi_\mathrm{S}$ is also observed. If we construct the polarization angle quantity from data integrated in all incident and emission angles, for the purpose of obtaining larger photon statistics, we observe $\Psi \approx 90^{\circ}$ at all energies (by $\lesssim 2^{\circ}$ variation in the total range), apart from a few remaining spectral lines that tend to manifest as single points in energy in our resolution around $90^{\circ}\pm 90^{\circ}$.

\begin{figure}
	\includegraphics[width=\columnwidth]{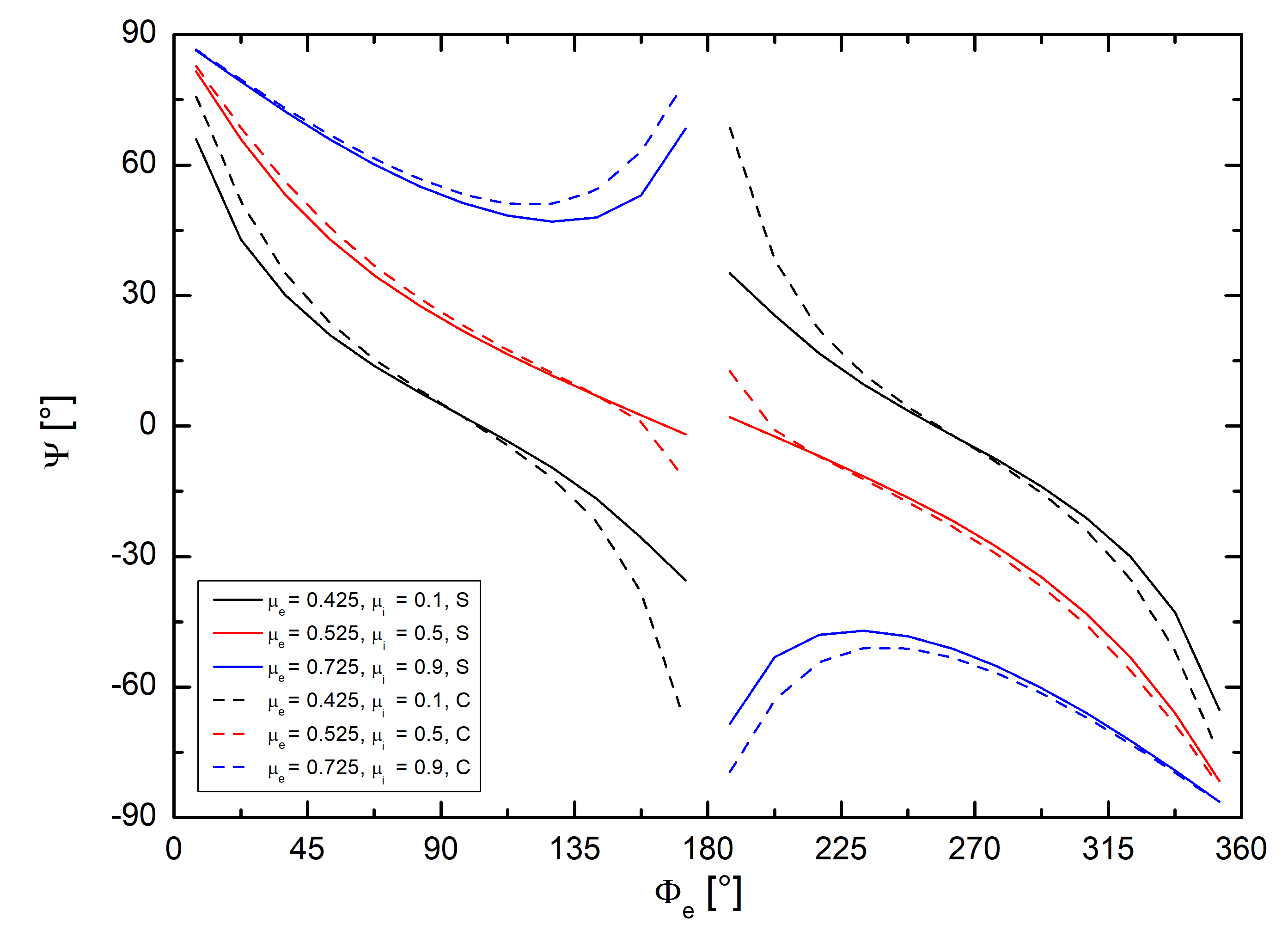}
	\caption{The polarisation angle versus azimuthal emission angle obtained by {\tt STOKES} (solid lines) integrated in $E$, $\Gamma$, and $\xi_\mathrm{S}$ for the unpolarized primary radiation and various $\mu_\mathrm{i}$, $\mu_\mathrm{e}$ combinations. Values for the single-scattering approximation by Chandrasekhar's formulae for unpolarized primary radiation in these geometries are shown (dashed lines).}
	%Corresponding figures for pure horizontal and diagonal primary polarisation are Figures \ref{fig:A19} and \ref{fig:A20} in the~Appendix.}
	\label{fig:psiphivar}
\end{figure}

Since the average value of $\Psi$ over the 10 neighbouring energy bins does not depend on $E$, $\Gamma$ and $\xi_\mathrm{S}$, it is also reasonable to integrate over these parameters in order to obtain better statistics on its geometrical behaviour without loss of generality. Similarly to Fig. \ref{fig:pphivar}, Fig. \ref{fig:psiphivar} displays $\Psi(\Phi_\mathrm{e})$ for various $\mu_\mathrm{i}$ and $\mu_\mathrm{e}$ values and in comparison to the same Chandrasekhar’s formulae for Rayleigh single-scattering of unpolarized incident radiation. Likewise, we obtained results close to the analytical approximation also for the cases of horizontal and diagonal primary polarization, which we omit here for brevity, and all of these figures can be explained by the role of scattering angle in polarization change upon single scattering, despite some distortion caused by multiple scattering and despite some numerical noise. {\tt STOKES} is again very precise in treating the scattering angle geometry with respect to the polarization vector located in the polarization plane, and the obtained polarization angle values follow the analytical behaviour quite nicely notwithstanding the fact that the $\Psi$ quantity, given its definition, in addition suffers from coordinate degeneracy.

\section{Discussion}\label{section4}

\subsection{Other spectral reflection models}

The spectral part of the {\tt TITAN} and {\tt STOKES} results obtained by us can be directly compared to similar attempts in the literature, unlike our polarization results, which are unique. The problematic of reflection spectra from accretion discs was already addressed by e.g. the computations with {\tt NOAR} \citep{Dumont2000} and {\tt PEXRAV} \citep{Magdziarz1995}. These models assumed a neutral disc for simplicity, which produces a considerable flux deficiency at soft X-rays that is not supported by the expected temperatures that impose complicated structures in terms of ionization. {\tt PEXRIV} \citep{Magdziarz1995} is another model already involving an ionized disc, but being too simple in assuming isothermal medium with a maximum temperature of $T = 10^6 \textrm{ K}$. Another attempt for a reflection code involving ionized disc was
completed by\cite{Ross1993, Ross1999, Ross2005} with the {\tt REFLIONX} models. The most recent major step further was achieved by \cite{Garcia2013}, where the ionization structure up to optical depth $\sim10$, was computed by the {\tt XSTAR} code \citep{Kallman2001}, and spectral reflection tables were subsequently computed using the Feautrier radiative transfer solving method \citep{Mihalas1978}, resulting in the {\tt XILLVER} tables \citep{Garcia2010, Garcia2011, Garcia2013}. The paper by \cite{Garcia2013} also introduces a discussion and comparison of these up-to-date most precise spectral reflection tables in X-rays with computations using the older {\tt REFLIONX} tables that utilize Fokker–Planck diffusion equation, including a modified Kompaneets operator \citep{Ross1978, Ross1979}, and that also already assume an ionized structure, but e.g. underestimate the amount of neutral atoms \citep{Garcia2013}.\footnote{All the differences between {\tt REFLIONX} and {\tt XILLVER} on the spectral output are still not fully understood \citep{Garcia2013}.}

In the next subsection, our contribution to this problem will be the presentation of a comparison of the combined {\tt TITAN} and {\tt STOKES} local spectral tables in the reflection scenario with the {\tt REFLIONX} tables and {\tt XILLVER} tables that are the most widely used models by the scientific community today and that greatly resemble our setup. In the future, we plan to compare our local results with the {\tt XILLVER} results in the global view, i.e. as if the inner accretion system was observed by a distant observer, including the primary source.

\subsection{Direct comparison of our spectral results with the {\tt XILLVER} and {\tt REFLIONX} tables}

\begin{figure}
	\includegraphics[width=\columnwidth]{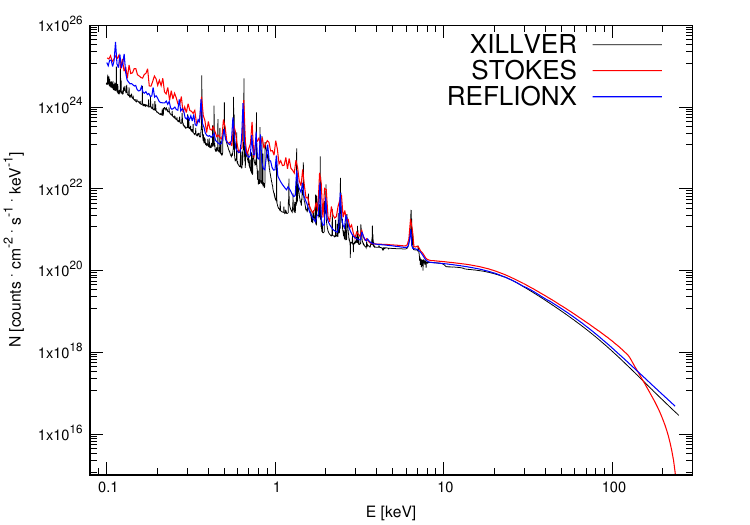}
	\caption{Comparison of the {\tt XILLVER} tables (black)  for $\xi_\mathrm{X} = 100$ and $\Gamma = 2.2$, and {\tt REFLIONX} tables (blue) for $\xi_\mathrm{R} = 100$ and $\Gamma = 2.2$ with the corresponding {\tt STOKES} spectral results (red) for $\xi_\mathrm{S} = 106.5$ and $\Gamma = 2.2$, all normalized accordingly.}
	\label{fig:xillvercomp}
\end{figure}

The {\tt XILLVER} reflection spectra, which are also available in the FITS format \citep{Garcia2013}, assume unpolarized primary radiation and are integrated in $\mu_\mathrm{i}$ and $\Phi_\mathrm{e}$ parameters. As spectral dependency on $\mu_\mathrm{e}$ in terms of shape and amplitude is not dramatic, it is reasonable to average {\tt XILLVER} tables in all $\mu_\mathrm{e}$ in order to obtain better statistics and confine to the average $\mu_\mathrm{e}=0.5$ value, which is also an average inclination angle for {\tt STOKES} in our mesh set-up. Comparison of the {\tt STOKES} spectral results under unpolarized incident illumination with {\tt XILLVER} tables is shown on Fig. \ref{fig:xillvercomp}. For this comparison, the raw {\tt STOKES} output $I$ was averaged and normalized such as
\begin{equation}\label{xilnorm}
	N(E) = \dfrac{\xi_\mathrm{S} n_\mathrm{H} \int_{E_\mathrm{min}}^{E_\mathrm{max}}E^{-\Gamma}dE}{B N_{\mu_\mathrm{e}} 4 \pi^2 e \times 10^{10}} \sum_{\mu_\mathrm{e},\mu_\mathrm{i},\Phi_\mathrm{e}} \mu_\mathrm{i} \Delta \mu_\mathrm{i} \dfrac{I(E; \mu_\mathrm{e}, \mu_\mathrm{i}, \Phi_\mathrm{e})}{N_\mathrm{tot}\Delta \mu_\mathrm{e} \Delta E} \textrm{ ,}
\end{equation}
where
\begin{equation*}
	B = E_\mathrm{X,c}^{2-\Gamma}[\Gamma_\mathrm{f}(2-\Gamma,\dfrac{E_\mathrm{X,0}}{E_\mathrm{X,c}})-\Gamma_\mathrm{f}(2-\Gamma,\dfrac{E_\mathrm{X,1}}{E_\mathrm{X,c}})],
\end{equation*}
and
\begin{equation*}
	\Gamma_\mathrm{f}(s,x) = \int_{x}^{\infty}t^{s-1}\mathrm{e}^{-t}dt
\end{equation*}
is the upper incomplete gamma function and $E_\mathrm{X,0} = 0.1 \textrm{ keV}$, $E_\mathrm{X,1} = 1000 \textrm{ keV}$, $E_\mathrm{X,c} = 300 \textrm{ keV}$ are values related to the used {\tt XILLVER} cut-offs. The normalization (\ref{xilnorm}) differs from (\ref{norm2}), because apart from the units and geometrical conventions, it is necessary to again account for alternative cut-offs used in {\tt XILLVER} tables as opposed to the {\tt STOKES} computations. Namely, the high-energy cut-off in {\tt {\tt XILLVER}} is exponential and was selected as $E_\mathrm{X,c} = 300 \textrm{ keV}$ ({\tt XILLVER} tables offer to choose $E_\mathrm{X,c}$ as a free model parameter) and low-energy cut-off is sharp at $E_\mathrm{X,0} = 0.1 \textrm{ keV}$. Hence, the primary radiation there follows equation
\begin{equation}
	N(E) = N_\mathrm{0,X}E^{-\Gamma}e^{-\frac{E}{E_\mathrm{X,c}}} \textrm{ ,}
\end{equation}
where $N_\mathrm{0,X}$ corresponds to $\xi_\mathrm{X}$, which we further use to denote the ionization parameter in {\tt XILLVER} grid, via
\begin{equation}
	\xi_\mathrm{X} = \dfrac{4 \pi N_\mathrm{0,X}}{n_\mathrm{H}} B \textrm{ .}
\end{equation}
In {\tt STOKES}, both cut-offs are sharp and primary radiation follows equation (\ref{primary}). In Fig. \ref{fig:xillvercomp}, the original {\tt XILLVER} tables are in addition divided by $\Delta E$ at each bin and multiplied by the necessary factor $\dfrac{10^{20}}{4 \pi}\mu_\mathrm{e}$ that accounts for FITS storage convention and geometrical definitions therein. Last but not least, we also multiplied the {\tt XILLVER} tables by $\xi_\mathrm{S}/\xi_\mathrm{X}$ factor to compensate for the differences in $\xi$, as this parameter does not have the same grid in both computations and directly determines the amplitude at all energies. In order to compare both models efficiently, we selected the closest $\xi_\mathrm{S}$ parameter in {\tt STOKES} tables to our $\xi_\mathrm{X}$ selection. {\tt XILLVER} uses similar grid in $\Gamma$ parameter and the same values could be used for comparison without a need of interpolation. The same holds for elemental abundance $A_\mathrm{Fe} = 1.0$ present in both tables, but we note that the solar chemical composition used in {\tt XILLVER} (from \citealt{Grevesse1998}) is severely outdated with respect to the one used in {\tt STOKES} (from \citealt{Asplund2005}), see the discussion in \citet{Grevesse2007}.

At the Compton hump energies ($\sim 20 \textrm{ keV}$), both spectra remarkably match with the adopted normalization. This holds similarly for other comparable ionization parameters and photon-indices, if displayed, as Fig. \ref{fig:xillvercomp} is only an example of such comparison. We stress this conclusion, since our tables were obtained by a completely different (Monte Carlo) method, while {\tt XILLVER} tables were obtained by simultaneously solving the equations of radiative transfer, energy balance, and ionization equilibrium (also in a Compton-thick, plane-parallel medium), using the Feautrier method \citep{Garcia2010, Garcia2011, Garcia2013}). Therefore, our local results are cross-checked in the hard X-rays before further use in, e.g. total spectral production of AGNs, or before further exploration of our technique.

Juxtaposition of models at these energies makes most sense, because $\xi$ variation and subsequent compensation for uneven $\xi$ grid does not permeate heavily at these energies (see e.g. Fig. \ref{fig:spectrumxi}), although the uneven grid in $\xi$ is believed to be the biggest issue in setting the normalization accordingly and inherently leaves discrepancies in amplitude. Also, line production based on different atomic data produced by {\tt TITAN} and {\tt XSTAR} is not present at this range.

In \cite{Garcia2013}, the strong influence of cut-off definition chosen in the primary power law on the spectral shape, which also has to be taken into account when comparing two different codes, was
pointed out. In particular, \cite{Garcia2013} illustrates the impact of different low-energy (continuous and broken power law) and high-energy (exponentials with different $E_\mathrm{X,c}$) cut-offs on the reflected local spectrum acquired in {\tt XILLVER}. This manifests itself especially at energies below 1 keV. The spectral differences due to alternative cut-off shapes and positions can reach one order of magnitude. The change in overall spectrum is stronger, if higher $\Gamma$ is used \citep[see figs 1 and 2 in][]{Garcia2013}. This is inherently connected to the temperature profile with optical depth $\tau$ inside the disc. For our case, the temperatures computed by {\tt TITAN} stay in between the profiles of {\tt XILLVER} (provided for similar parametric set-up in \cite{Garcia2013}) until  $\tau \sim 1$ but in deeper layers the disc remain about an order of magnitude hotter in {\tt TITAN}. Also, our computations ended at layers with $\tau \sim 7$, which is less by 3 than the depths used in radiative transfer computations of {\tt XILLVER} tables, which may be another reason behind the discrepancies below 1 keV. The fact that {\tt XILLVER} and {\tt STOKES} results vary in some line features and their strength is not surprising, as different atomic data are used, our computations include $\sim 4000$ atomic transitions, which is nearly half of the spectral lines that are considered in {\tt XSTAR}, and for constructing {\tt XILLVER} models calculations of level populations, temperature, total opacity, and emissivity are performed by {\tt XSTAR}, not {\tt TITAN}.

In the attempts of comparing {\tt XILLVER} results to {\tt REFLIONX}, {\tt PEXRAV} or {\tt PEXRIV} tables in \cite{Garcia2013}, the differences also mostly appeared in the soft energy range. To contribute to this discussion, we displayed the corresponding {\tt REFLIONX} computations along our comparisons between {\tt STOKES} and {\tt XILLVER}. The {\tt REFLIONX} tables \citep{Ross1993, Ross1999, Ross2005} are stored in the same FITS format as the {\tt XILLVER} tables and contain the same high- and low-energy cut-offs. Thus, we again choose the closest ionization parameter $\xi_\textrm{R}$ to our $\xi_\textrm{X}$ and $\xi_\textrm{S}$ selection, the same power-law index $\Gamma$ and the same elemental abundance $A_\mathrm{Fe} = 1.0$ \citep[but now with a solar chemical composition from][]{Morrison1983} and we normalize {\tt REFLIONX} in the same way as {\tt XILLVER} with an additional factor of 2, which accounts for the fact that the {\tt REFLIONX} tables are already averaged in $\mu_\textrm{e}$. We conclude that at soft X-rays, where most discrepancies appear, the STOKES models are typically closer to the {\tt REFLIONX} computations, rather than the {\tt XILLVER} computations. An example is Fig. \ref{fig:xillvercomp}, where the {\tt REFLIONX} spectrum is added. A more detailed comparison of the soft X-ray results obtained by {\tt STOKES} and other computational methods is required to understand the apparent differences in the soft X-ray range. We plan to do this in future work. Here, we focus on the X-ray polarization results that are not sensitive to the spectral discrepancies around 1 keV.

Weakness of our spectral model compared to the {\tt XILLVER} computations is the neglect of Compton up-scattering in the disc matter in {\tt STOKES} code. We may also include an extra thermal component of the X-ray radiation originating from the disc and synchrotron processes to our model in the future that would make the {\tt STOKES} X-ray tables more applicable to XRBs and more precise in the soft X-rays. A floating grid in primary cut-off boundaries, non-constant density, or variable elemental abundance is also something that can be implemented in the future versions of our models and so far has not been tested due to large computational times. Our primary objective was to timely deliver the first very realistic polarization computations in the reflection scenario for the upcoming X-ray polarimetric missions that operate at hard X-rays above 2 keV.

Because {\tt STOKES} accounts for multiple scattering, it is almost impossible to give a quantitative insight of the importance of inverse Compton scattering and synchrotron emission. Howver, qualitatively, inverse Compton scattering will likely smooth the high-energy tail of the Compton hump due to energy shifts \citep{Garcia2020} and decrease a little bit the polarization degree of our current results. In general, inverse Compton scatterings tend to reduce the polarization degree of emission; multiple Compton scatterings do so even more \citep{Krawczynski2011}. Unpolarized synchrotron photons give rise to synchrotron self-Compton emission with vanishing or very small (<5 per cent) polarization degrees for all but very low Lorentz factors \citep{Celotti1994, Begelman1987}. What is interesting is that the inverse Compton emission should track the polarization degree and polarization direction of the synchrotron emission. All of this is of prime importance for radio-loud objects. Our paper focuses on radio-quiet sources, so at first order we can safely neglect those effects (synchrotron emission in particular, which is in addition more relevant to optically thin media, rather than to optically thick standard disc), but we admit that inverse Compton scattering should be included in future simulations.

An advantage of our spectra over the {\tt XILLVER} tables is the model dependency on $\mu_\mathrm{i}$ and $\Phi_\mathrm{e}$ angles, which is more demanding on the total number of photons $N_\mathrm{tot}$ for the Monte Carlo method on one hand, but allows higher geometrical precision and angular checks that involve macroscopic analysis of scattering processes in the disc medium. Angular precision in the scattering simulation is especially crucial for the polarization outcome, as the creation/alteration of polarization is undoubtedly connected to (a)symmetries in the origin of radiation and accurate geometrical shapes. So far we have illuminated our disc slab with isotropic radiation for the pre-computations with {\tt TITAN} but here we could also acquire higher precision in the future with incident inclination-dependent computations of ionization structure. An advantage of our new model worth mentioning is that incident arbitrary polarization of the primary radiation is now enabled, which is of course more important in the context of polarimetry, as it was already discussed in Section \ref{section3}. In the future, we plan to develop tables with a fixed grid in ionization parameter that would also have a much higher resolution in energy. Such tables would enable further discussion upon spectral line features and disc structure modelling, although they might not be of any use to the low-resolution and photon-demanding X-ray polarimetry.

\subsection{Further validation of our polarization results and implications}

Despite some weaknesses of our current model, the spectral comparisons already validate a discussion of polarization of the reflected emission obtained uniquely by {\tt STOKES} in the energy ranges of the forthcoming polarimetric missions IXPE or eXTP ($\gtrsim 2 \textrm{ keV}$) and further in the hard X-rays. Moreover, Fig. \ref{fig:pxivar} showed generally higher predicted polarization fraction above 2 keV than in the soft X-rays due to the polarization induced by Compton down-scattering in the disc. Although this is promising for targeting AGN and XRB sources by the IXPE or eXTP missions, the total polarization measured by a distant observer is expected to be lower due to the impact of integration over the disc \citep[even though in some specific configurations the polarisation degree can be \textit{magnified} due to general-relativistic effects, see][]{Dovciak2011} and the impact of direct primary radiation \citep[expected to be mildly  polarized, $\lesssim 10$ per cent, see e.g.][]{Tamborra2018, Beheshtipour2017, BeheshtipourThesis} reaching the observer.

In Fig. \ref{fig:pphivar}, we already showed that the polarization degree predicted by our semirealistic simulation is lower than the analytical formulae forecast due the presence of multiple scattering and general crudeness of the Chandrasekhar’s approximation. The Chandrasekhar’s formulae represent a single-scattering approximation of elastic Rayleigh scattering without the ability to reconstruct characteristic Compton recoil around 20 keV or the effects caused by spectral lines. None the less, they confirm the output of our simulation with respect to basic angular parameters and this approximation may also confirm basic energy dependence of the polarization quantities, especially the expected tendency of $p$ to rise towards harder X-rays, in \textit{global} models once we apply them to the new local reflection tables \cite[see e.g.][]{Dovciak2011}.

\section{Conclusions}\label{section5}

In this paper, we discussed a method to simulate spectrum and polarization of locally reflected X-ray radiation from BH accretion discs. We chose a power-law primary source that represents a hot
gaseous corona situated above a semi-infinite disc slab. The ionization structure of the disc was pre-computed by a radiative transfer code {\tt TITAN} \citep{Dumont2003}, suitable for ionized optically thick media. The low-resolution reflection tables comprising all linear ($I$, $Q$ and $U$) Stokes parameters were computed by a Monte Carlo code {\tt STOKES} \citep{Goosmann2007,Marin2012,Marin2015,Marin2018} that incorporates important X-ray line and continuum mechanisms, including the characteristic multiple Compton down-scattering of X-ray photons in the partially ionized disc medium. We computed the energy-dependent X-ray outcome, i.e. the Stokes parameters $I$, $Q$ and $U$, for three arbitrary polarization states of the primary radiation, numerous ionization parameters $\xi_\mathrm{S}$, photon-indices $\Gamma$, incident inclinations $\mu_\mathrm{i}$, and emission angles $\mu_\mathrm{e}$ and $\Phi_\mathrm{e}$. The results are made available to the reader in a compact form of the FITS format in the attachment of this paper. The new simulation can be methodically compared to spectral reflection models of similar kind appearing in the literature and it represents a unique and first attempt to numerically compute X-ray polarization in the local reflection scenario.

We showed that with all the model assumptions and physical processes incorporated in {\tt TITAN} and {\tt STOKES} we were able to produce X-ray spectra possessing the most common reflection features supported by up-to-date theoretical, numerical, and observational information on radio-quiet X-ray compact accreting sources, such as the Compton hump at around 20 keV, a sharp decline at $E > 10^2 \textrm{ keV}$, an excess at soft energies, a forest of lines around 1 keV, and the most prominent Fe K$\alpha$ line at $6.4$--$7 \textrm{ keV}$. The ionization parameter, defined by (\ref{xi}), correctly enforces stronger reflection with larger illumination and the reflected spectrum starts to resemble the original power-law shape at high $\xi_\mathrm{S}$ values. We found that extreme initial polarization of the primary radiation may affect significantly the spectral output in selected geometries, but if integrated over the incident and emission angles, which can imitate a subsequent integration over the disc for total spectral production (as the source would be observed from infinity), we acquired almost identical spectra.

We managed to validate our method by successful normalization and comparison to the up-to-date most widely used reflection spectral tables obtained by {\tt XSTAR} called {\tt XILLVER} \citep{Kallman2001,Garcia2010, Garcia2011,Garcia2013}, as well as the older {\tt REFLIONX} reflection spectra \citep{Ross1993, Ross1999, Ross2005}. The comparison was achieved with similar model parameters and complete angular integration of these models. The spectra appeared very similar above 3 keV, which is positive for the 2--12 keV energy range that will be explored by the upcoming polarimetric missions, for which we primarily develop our spectropolarimetric models. The remaining discrepancies, especially in the soft X-rays, may lie in the uneven $\xi$ grid between the tables, different atomic data, adopted optical depths, selected cut-offs of the primary radiation, or the computational methods themselves, especially at the disc structure numerical stage. In the future, we may improve our {\tt STOKES} local simulation by incorporating Compton up-scattering, and disc black-body radiation and include more flexibility in the disc’s density,
elemental abundance, or primary cut-off selection.

From our results we may already discuss the polarization properties of locally reflected X-ray radiation. We showed that reflection can locally induce large polarization degrees despite multiple
scattering and other depolarization factors. Comparing our models to the available Chandrasekhar’s analytical approximation of single scattering in a slab region \citep{Chandrasekhar1960}, we confirmed the expected behaviour of polarization degree and angle with respect to the incident and emission angles in the local co-moving frame. We confirmed the relative importance of the initial polarization state of the primary radiation that imprints the polarized outputs. In the crude resolution of $\Delta \log E = 0.1$, the local polarization angle remains independent of $E$, $\Gamma$, and $\xi_\mathrm{S}$ and variations of the local polarization degree with $\xi_\mathrm{S}$ are prominent in the whole energy range. In the soft part of the spectrum, the polarization degree increases with ionization to the detriment of general depolarization caused by spectral lines, while in the medium-to-hard energy range (between 2 and 30 keV) it decreases with ionization, probably due to lower absorption \citep[larger absorption induces also larger polarization degree in the X-ray thermal emission in case of accretion discs in X-ray binaries studied by][] {Taverna2020b}. The fact that the polarization degree of the reflected X-ray emission can be, locally, quite large (i.e. tens of per cents) between 2 and 12 keV, is a good sign for the forthcoming X-ray polarimetric missions that operate at these frequencies.

With these low-resolution tables, we may already produce some total spectra and discuss the global spectropolarimetric properties of BH accreting sources, as well as compute more realistic estimates of observational times needed for the forthcoming X-ray polarimetric missions. This will be properly addressed in our future works. It has been argued already in the introduction that reflection models are an important part of consistent modeling of compact radio-quiet accreting sources. We believe that the presented approach may bring fresh insights into the X-ray spectral modelling of inner accretion phenomena and that the polarization tables achieved in this way will later become a valuable component for the total emission models and for data fitting of AGN and XRB sources that will soon be needed for the X-ray polarimetric mission IXPE \citep{Weisskopf2013, Weisskopf2016}, due to be launched in 2021, and that may become valuable in terms of constraining inclination and orientation of real sources, BH spins, coronal morphologies or disc’s structures.

\section*{Acknowledgements}

MD thanks for the support from the GACR project 21-06825X and the institutional support from RVO:67985815. JP acknowledges financial support from the Charles University, project GA UK no. 174121, and from the Barrande Fellowship Programme of the Czech and French governments. AR was supported by Polish National Science Center grant no. 2015/17/B/ST9/03422.

%%%%%%%%%%%%%%%%%%%%%%%%%%%%%%%%%%%%%%%%%%%%%%%%%%
\section*{Data Availability}

The model reflection tables underlying this article (created and compiled by authors of this article) are available in the FITS format in the Figshare Repository at \url{https://doi.org/10.6084/m9.figshare.16726207}.



\bibliographystyle{mnras}
\bibliography{example} % if your bibtex file is called example.bib

%\begin{thebibliography}{99}
%\end{thebibliography}

%%%%%%%%%%%%%%%%%%%%%%%%%%%%%%%%%%%%%%%%%%%%%%%%%%

%%%%%%%%%%%%%%%%% APPENDICES %%%%%%%%%%%%%%%%%%%%%

%\appendix


%%%%%%%%%%%%%%%%%%%%%%%%%%%%%%%%%%%%%%%%%%%%%%%%%%

% Don't change these lines
\bsp	% typesetting comment
\label{lastpage}
\end{document}